\documentclass{PoS}
\usepackage{subfigure}
\usepackage{amsmath}

\title{Axion as a non-WIMP dark matter candidate}

\ShortTitle{Axion as a non-WIMP dark matter candidate}

\author{\speaker{Ken'ichi Saikawa}\\
      Deutsches Elektronen-Synchrotron DESY\\
      Notkestrasse 85, D-22607 Hamburg, Germany\\
      E-mail: \email{kenichi.saikawa@desy.de}}

\abstract{
The axion arises in well-motivated extensions of the Standard Model of particle physics and is regarded as 
an alternative to the weakly interacting massive particle paradigm to explain the nature of dark matter.
In this contribution, we review theoretical aspects of dark matter axions, particularly focusing on recent developments in
the estimation of their relic abundance.
A closer look at their non-thermal production mechanisms in the early universe reveals 
the possibility of explaining the observed dark matter abundance in various mass ranges.
The mass ranges predicted in various cosmological scenarios are briefly summarized.
}

\FullConference{The European Physical Society Conference on High Energy Physics\\
		 5-12 July\\
		 Venice, Italy}

\begin{document}

\section{Introduction}
Dark matter is one of the largest outstanding issues of modern particle physics and cosmology.
Although we have several observational evidences of its existence, its microscopic properties 
still remain unknown.
At this stage, we can only enumerate various necessary conditions for a dark matter candidate:
it should be cold, neutral, abundant enough to match the appropriate relic density, and compatible with
all astrophysical constraints [see, {\it e.g.}, Ref.~\cite{Taoso:2007qk}]. 
The Weakly Interacting Massive Particles (WIMPs) motivated by new physics at TeV energy scale
is nicely fitted to these requirements,
but there is plenty of room for consideration of non-WIMP candidates.
Moreover, non-WIMP dark matters are now getting more attention
because of the fact that the LHC experiment does not find new physics and that
recent direct detection experiments put severe constraints on the WIMP parameter space.

Among a lot of possibilities, 
the axion can be regarded as a leading candidate of non-WIMP dark matter,
as it arises in well-motivated extensions of the Standard Model~\cite{Peccei:1977hh} and has a rich phenomenology. 
In this contribution, we discuss theoretical aspects of dark matter axions,
particularly highlighting recent developments in the estimation of their relic abundance.

\section{QCD axion as a dark matter candidate}
The existence of the axion was postulated in the Peccei-Quinn (PQ) mechanism~\cite{Peccei:1977hh},
which was introduced as a solution to the strong CP problem of quantum chromodynamics (QCD).
In the PQ theory, we assume the presence of a global $U(1)_{\rm PQ}$ symmetry, which has to be spontaneously broken
at some high energy scale. The axion can be identified as a (pseudo) Nambu-Goldstone boson
associated with the spontaneous breaking of the $U(1)_{\rm PQ}$ symmetry~\cite{Weinberg:1977ma}.
This global symmetry is explicitly broken due to the QCD anomaly, which induces a mass of the axion~\cite{diCortona:2015ldu},
\begin{equation}
m_a = 57.0(7)\,\mathrm{\mu eV}\left(\frac{10^{11}\,\mathrm{GeV}}{f_a}\right)\,,
\label{axion_mass}
\end{equation}
where $f_a$ is associated with the PQ symmetry breaking scale and called the axion decay constant.
Although the axion has a tiny mass, it behaves as a cold matter in the universe since it can be produced non-thermally~\cite{Preskill:1982cy}.
Furthermore, its interactions with ordinary matter are inversely proportional to $f_a$, and they are highly suppressed if
$f_a$ is sufficiently large. These features make the axion suitable for explaining the nature of dark matter.

In the axion dark matter scenario, there are subtle issues in the estimation of its relic abundance.
We naively expect that it should be possible to determine the typical mass of the axion dark matter
via Eq.~\eqref{axion_mass} 
provided that we correctly estimate the present axion abundance $\Omega_a$
(the ratio of the energy density of axions to the critical density of the universe) as a function of $f_a$.
However, this procedure is not so straightforward due to the fact that
there are several theoretical uncertainties in the estimation of $\Omega_a$.
The concerns are twofold: (1) We need to know the finite temperature effective potential
that describes the evolution of the axion field during the epoch of the QCD phase transition.
(2) Furthermore, the estimation strongly depends on the assumption about the early history of the universe.

The effective potential for the axion field $a(x)$ at finite temperature $T$ can be modeled by the following form,
\begin{equation}
V(a,T) = \chi(T)\left[1-\cos\left(N_{\rm DW}\frac{a}{v_{\rm PQ}}\right)\right]\,, \label{axion_potential}
\end{equation}
where $v_{\rm PQ} = f_a N_{\rm DW}$ is the PQ symmetry breaking scale,
and $N_{\rm DW}$ is a positive integer called the domain wall number.
$\chi(T)$ is the topological susceptibility, which determines the temperature dependence
of the axion mass, $m_a(T)^2 = \chi(T)/f_a^2$.
It is believed that $\chi(T)$ rapidly grows towards its zero-temperature value
around the epoch of the QCD phase transition.
Since the axion field starts to evolve towards the minimum of the effective potential~\eqref{axion_potential}
when the Hubble parameter $H$ becomes comparable with $m_a(T)$,
it is important to know the temperature dependence of $\chi(T)$ to
follow the evolution of the axion field precisely.
The estimation of $\chi(T)$ is a tough problem, as we must analyze the non-perturbative effects in QCD.
Here we adopt the recent result of lattice calculations in full QCD~\cite{Borsanyi:2016ksw},
which can be fitted to the power law $\chi(T) \propto T^{-n}$ with $n=8.13$ at high temperatures
and is compatible with the estimate based on the dilute instanton gas approximation.

In addition to the issue about the effective potential around the epoch of the QCD phase transition,
it is also important to take account of the evolution of the axion field throughout the whole history of the universe.
In particular, we can consider two different possibilities according to whether the PQ symmetry has been broken before/during
inflation or not. If the PQ symmetry has been broken before/during inflation and never been restored (the \emph{pre-inflatuonary} PQ symmetry breaking scenario),
a region in which the axion field takes a certain value $a_i$ is rapidly expanded 
during the inflationary epoch. As a consequence, the axion field has a single uniform initial value $a_i$ within the observable universe.
On the other hand, if the PQ symmetry has been restored and broken after inflation (the \emph{post-inflatuonary} PQ symmetry breaking scenario), 
the axion field takes different values in different spatial regions.
In this case, we need to carefully investigate how such inhomogeneous field configurations evolve in the early universe.

One of the most important features of the post-inflationary PQ symmetry breaking scenario is that it predicts the formation of
topological defects in the early universe, and that axions can be produced from them~\cite{Davis:1986xc}.
In axion cosmology, two kinds of topological defects are relevant: 
One is the string, which is formed when the PQ symmetry is spontaneously broken.
The other is the domain wall, which is associated with dynamics of the axion field due to the potential~\eqref{axion_potential}
during and after the epoch of the QCD phase transition~\cite{Sikivie:1982qv}. 
Around the epoch of the QCD phase transition, the axion field settles down at one of $N_{\rm DW}$ minima of the potential~\eqref{axion_potential}
in each spatial region, which creates $N_{\rm DW}$ different domains within the observable universe.
Domain walls are sheet-like objects located around the boundaries of such domains.

Since the axion field $a$ must vary from $0$ to $2\pi v_{\rm PQ}$ around the core of the string,
strings are always attached by $N_{\rm DW}$ domain walls. The subsequent evolution of such string-wall systems
is different according to the value of $N_{\rm DW}$, which is determined from the QCD anomaly coefficient and hence depends on underlying particle physics models.
If $N_{\rm DW} = 1$, the string is attached by one domain wall, and the string-wall systems collapse soon after the formation due to the tension of domain walls.
On the other hand, if $N_{\rm DW} > 1$, the string-wall systems are stable, since the strings are pulled in $N_{\rm DW}$ different directions.
The existence of such stable domain walls is in serious contradiction with the standard cosmology~\cite{Zeldovich:1974uw}.
However, there is a possibility to avoid this domain wall problem if there exists an explicit symmetry breaking operator in addition to the potential~\eqref{axion_potential}~\cite{Sikivie:1982qv}.
In the presence of the explicit symmetry breaking operator, string-wall systems become unstable,
and their late-time collapse also leads to the production of cold axions.

\section{Relic axion abundance and predicted mass ranges}
According to the history of the universe, we can consider three different possibilities:
(1) the pre-inflationary PQ symmetry breaking scenario, (2) the post-inflationary PQ symmetry breaking scenario with $N_{\rm DW} = 1$,
and (3) the post-inflationary PQ symmetry breaking scenario with $N_{\rm DW} >1$.
In the following, we discuss how the relic axion abundance is estimated in each scenario.

\subsection{Pre-inflationary PQ symmetry breaking scenario}
If the PQ symmetry has never been restored after inflation, the axion field takes an universal initial value
parameterized by $\theta_i \equiv a_i/f_a$. The production of dark matter axions in this case is well described by so called the vacuum
realignment mechanism~\cite{Preskill:1982cy}.
In this scenario, the axion field starts to oscillate coherently around the minimum of the potential~\eqref{axion_potential}
when the axion mass $m_a(T)$ becomes comparable with the Hubble parameter $H$, and the axion number is fixed around that epoch.
The relic axion abundance in the regime $|\theta_i|\ll \pi$ is given by~\cite{Ballesteros:2016xej}
\begin{equation}
\Omega_a h^2 \approx 0.35 \left(\frac{\theta_i}{0.001}\right)^2\times
\left\{
\begin{array}{ll}
\left(\frac{f_a}{3\times 10^{17}\,\mathrm{GeV}}\right)^{1.17} & \quad \text{for}\quad f_a\lesssim 3\times 10^{17}\,\mathrm{GeV}\\
\left(\frac{f_a}{3\times 10^{17}\,\mathrm{GeV}}\right)^{1.54} & \quad \text{for}\quad f_a\gtrsim 3\times 10^{17}\,\mathrm{GeV}
\end{array}
\right.\,,
\label{omega_a_pre}
\end{equation}
where $h=H_0/100\,\mathrm{km}\cdot\mathrm{sec}^{-1}\mathrm{Mpc}^{-1}$ is the reduced Hubble constant.
Note that the relic abundance depends not only on $f_a$ but also on the initial misalignment angle $\theta_i$.
Because of this fact, the axion can be the main constituent of dark matter in a broad range of $f_a$,
but we need to introduce a tuning $\theta_i \to 0$ for larger values of $f_a$ (or lower values of $m_a$). It is also possible to explain
the observed dark matter abundance in a higher mass range $m_a\lesssim 1\,\mathrm{meV}$ if we allow a tuning $\theta \to \pi$~\cite{Wantz:2009it}.

In the pre-inflationary PQ symmetry breaking scenario, quantum fluctuations of the axion field during inflation lead to large scale
isocurvature perturbations, and the null detection of such isocurvature perturbations in the observations of cosmic microwave background 
anisotropies puts stringent constraints on the parameter space~\cite{Wantz:2009it,Hertzberg:2008wr}.
In particular, this scenario can be excluded if the inflationary energy scale is sufficiently high.

\subsection{Post-inflationary PQ symmetry breaking scenario: models with $N_{\rm DW}=1$}
\begin{figure}
\centering
$\begin{array}{cc}
\subfigure{
\includegraphics[width=0.38\textwidth]{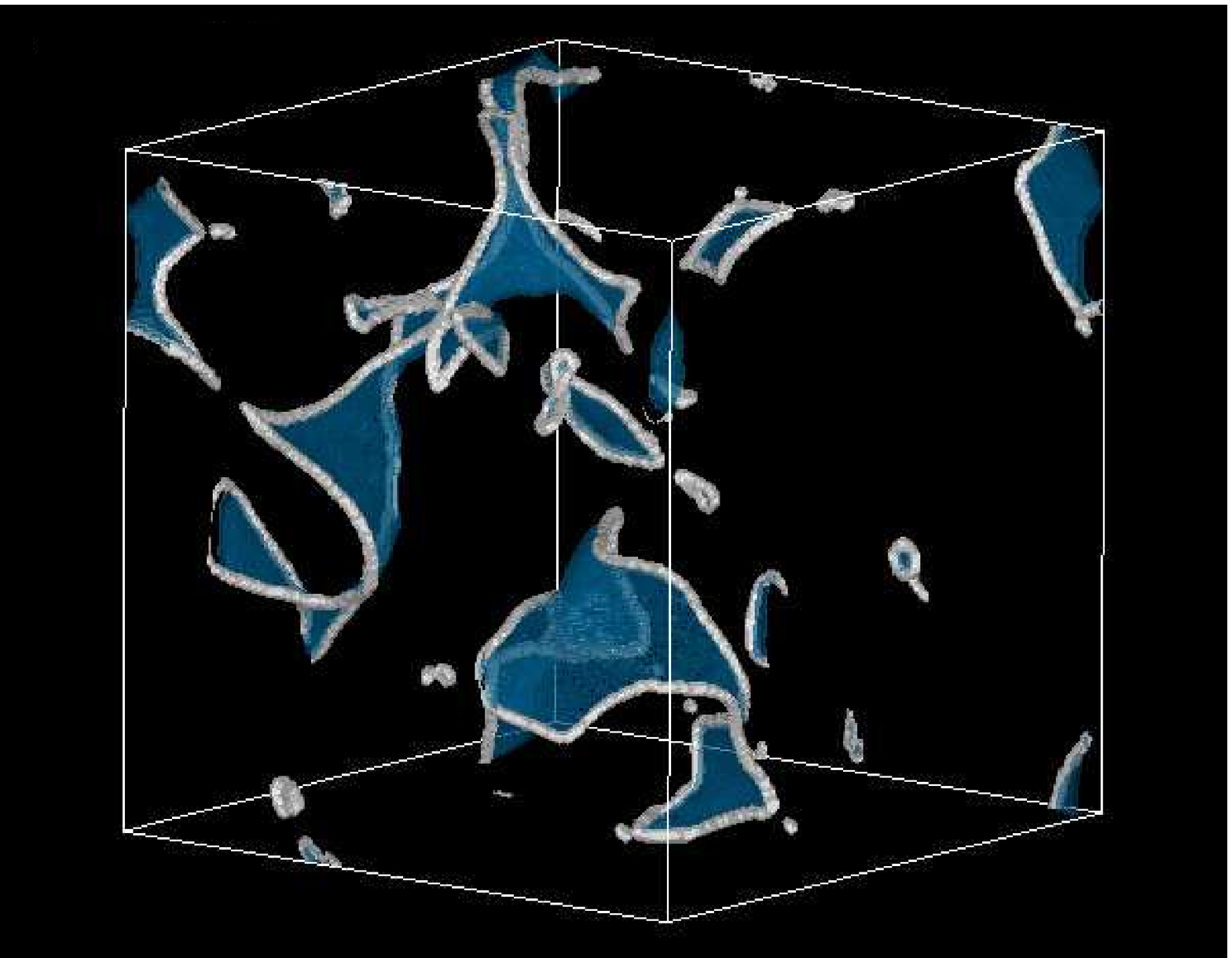}}
\hspace{25pt}
\subfigure{
\includegraphics[width=0.39425\textwidth]{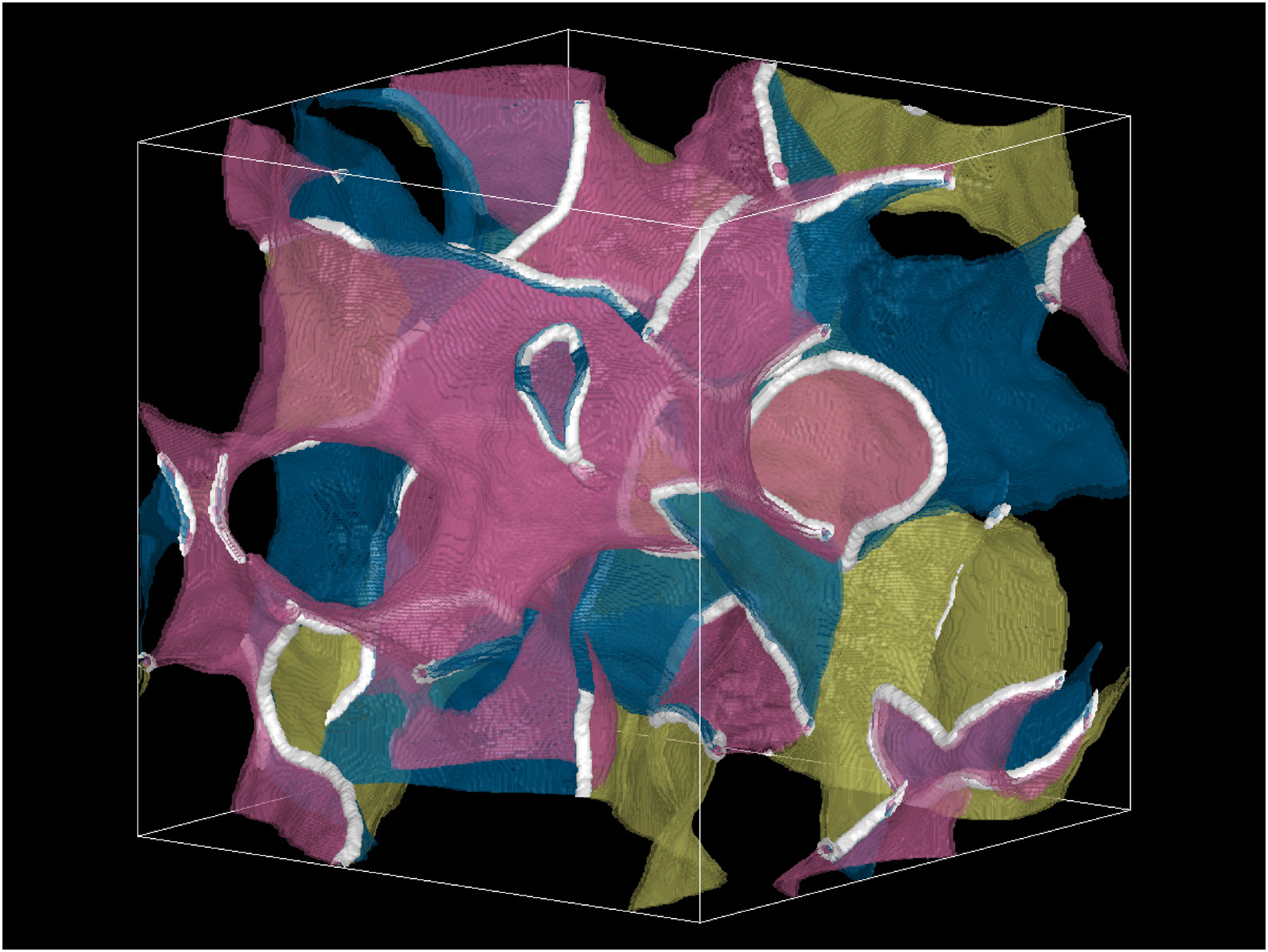}}
\end{array}$
\caption{Visualizations of numerical simulations of string-wall systems
for models with $N_{\rm DW}=1$ (left)~\cite{Hiramatsu:2012gg} and those with $N_{\rm DW} = 3$ (right)~\cite{Hiramatsu:2012sc}.
White regions correspond to the locations of the core of strings, and colored surfaces correspond to those of the center of domain walls.}
\label{fig1}
\end{figure}

In the post-inflationary PQ symmetry breaking scenario, axions can be produced from the decay of string-wall systems,
and we must take account of such contributions quantitatively.
Let us first consider the case with $N_{\rm DW} = 1$, in which there is no domain wall problem and the collapse of the string-wall systems
occurs around the epoch of the QCD phase transition.
The production of axions from the collapse of such short-lived string-wall systems was investigated
by using field theoretic lattice simulations~\cite{Hiramatsu:2012gg,Kawasaki:2014sqa},
whose visualization is shown in the left panel of Fig.~\ref{fig1}.
The results of the simulations imply that axions radiated from the string-wall systems are mildly relativistic,
and that the observed dark matter abundance can be explained in the mass range $m_a\approx (0.6$--$1.5)\times 10^{-4}\,\mathrm{eV}$,
which is higher than a naive estimate based on the realignment mechanism [{\it i.e.} averaging over $-\pi<\theta_i\le \pi$ in Eq.~\eqref{omega_a_pre}].

Recently, it was pointed out that the prediction for the axion dark matter mass
is potentially affected by the fact that strings have large tension,
which cannot be realized in the conventional simulation method~\cite{Fleury:2015aca}.
A modified method including the effects of large string tension was proposed in Ref.~\cite{Klaer:2017qhr}, which results in
a smaller value of the axion dark matter mass, $m_a = (2.62\pm 0.34)\times 10^{-5}\,\mathrm{eV}$.
Remarkably, the results of the modified simulations showed that the axion production is less efficient than
a naive estimate based on the realignment mechanism, despite of the fact that strings become denser than 
those observed in the conventional field theoretic simulations.
This fact implies that more energetic axions are produced, and that dynamics at smaller scales might be more important.
More detailed studies are required to confirm whether such shorter distance physics has a large impact on
the axion production efficiency.

\subsection{Post-inflationary PQ symmetry breaking scenario: models with $N_{\rm DW}>1$}
The domain wall problem is serious for models with $N_{\rm DW} > 1$, but it can be avoided if there exists
an operator which explicitly breaks the $U(1)_{\rm PQ}$ symmetry.
For instance, we can consider the following Planck-suppressed operators~\cite{Ringwald:2015dsf},
\begin{equation}
\mathcal{L} \supset g\frac{\Phi^N}{M_{\rm Pl}^{N-4}}+\mathrm{h.c.}\,, \label{symmetry_breaking_operator}
\end{equation}
where $\Phi$ is a complex gauge singlet scalar field that spontaneously breaks the global $U(1)_{\rm PQ}$ symmetry,
$M_{\rm Pl}$ is the reduced Planck mass, $N$ is an integer, and $g$ is a complex parameter.
These operators give rise to an additional term $\Delta V$ in the low energy effective potential for the axion field, which lifts degenerate minima of
the QCD potential~\eqref{axion_potential}, making domain walls unstable.
The lifetime of domain walls is determined by the height of the symmetry breaking term $\Delta V$, or the energy difference between quasi-degenerate vacua. 
Production of axions from such long-lived domain walls was investigated based on
the field theoretic lattice simulations~\cite{Kawasaki:2014sqa,Hiramatsu:2012sc},
whose visualization is shown in the right panel of Fig.~\ref{fig1}.
If $\Delta V$ is small enough, domain walls have lived for a long time and emitted a lot of axions, which tend to overclose the universe.
On the other hand, if $\Delta V$ is sufficiently large, it shifts the location of the minimum of the axion effective potential and leads to
large CP violation, spoiling the axionic solution of the strong CP problem.
In order to avoid these drawbacks, non-trivial conditions on the symmetry breaking parameters are required.

The detailed investigation of the parameter space showed
that there exits a loophole if the order of the operators~\eqref{symmetry_breaking_operator} is $N = 9$ or $10$~\cite{Ringwald:2015dsf}.
In such cases, the axion can explain the observed dark matter abundance in higher mass ranges, 
$5.6\times 10^{-4}\,\mathrm{eV} \lesssim m_a \lesssim 1.3\times 10^{-1}\,\mathrm{eV}$ (for $N_{\rm DW} = 6$),
if we allow a mild tuning of the symmetry breaking parameter $g$.
Intriguingly, such higher mass ranges are compatible with those preferred by 
stellar cooling anomaly observations~\cite{Giannotti:2017hny}.

\section{Conclusions}
\begin{figure}
\begin{center}
\includegraphics[width=0.95\textwidth]{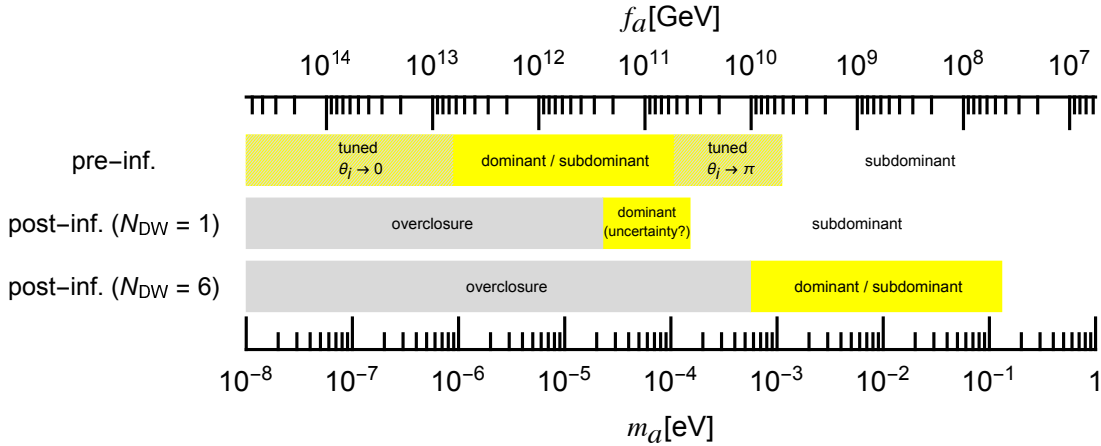}
\end{center}
\vspace{-5mm}
\caption{Predictions for the axion dark matter mass $m_a$ or the decay constant $f_a$ in the pre-inflationary PQ symmetry breaking scenario (first line),
the post-inflationary PQ symmetry breaking scenario with $N_{\rm DW} = 1$ (second line), and that with $N_{\rm DW} = 6$ (third line).
The yellow regions correspond to the mass ranges in which the axion can be the main constituent of dark matter.
The gray regions are excluded since the relic axion abundance exceeds the observed dark matter abundance.
The gray hatched regions correspond to the mass ranges in which more than 10\% tuning of $\theta_i$ is required in order to explain the observed dark matter abundance.
Here we give a conservative estimate of uncertainty in the axion dark matter mass for the models with $N_{\rm DW} = 1$, taking account of the difference
between the results obtained from the conventional simulation method~\cite{Hiramatsu:2012gg,Kawasaki:2014sqa} and those obtained from the modified simulation method~\cite{Klaer:2017qhr}.}
\label{fig3}
\end{figure}

The axion is a well-motivated hypothetical particle as it provides a solution to the strong CP problem
and can be a good candidate of non-WIMP dark matter.
The prediction for the axion dark matter strongly depends on the early history of the universe and hence the underlying particle physics models.
The mass ranges predicted in various cosmological scenarios are summarized in Fig.~\ref{fig3}.
Recently, a lot of new experimental projects are proposed, which enables us to investigate the properties of
the axion in the relevant parameter ranges [see, {\it e.g.}, Ref.~\cite{Graham:2015ouw}].
Discovery of the axion in such future experimental searches would bring about a tremendous development not only in dark matter physics but also in 
cosmology and fundamental physics.

\acknowledgments
We would like to thank
Maurizio Giannotti,
Takashi Hiramatsu,
Igor G. Irastorza,
Masahiro Kawasaki,
Javier Redondo,
Andreas Ringwald, and
Toyokazu Sekiguchi
for collaborations and discussions on the topics presented here.

\end{document}